\documentclass[aps,floatfix,preprintnumbers,amsmath,amssymb,amsfonts,final, sort, compress,12pt]
 {revtex4-1}

\usepackage{graphicx,color,natbib}

\begin{document}

\title{Statistical Mechanics Model for Protein Folding}

\author{Alexander V. Yakubovich$^*$, Andrey V. Solov'yov and Walter Greiner}

\affiliation{Frankfurt Institute for Advanced Studies, Goethe University,\\
60438, Ruth-Moufang Str. 1, Frankfurt am Main, Germany\\
$^*$E-mail: yakubovich@fias.uni-frankfurt.de}

\begin{abstract}
We present a novel statistical mechanics formalism for the theoretical description
of the process of protein folding$\leftrightarrow$unfolding transition in water
environment. The formalism is based on the construction
of the partition function of a protein obeying two-stage-like folding kinetics.
Using the statistical mechanics model of solvation of
hydrophobic hydrocarbons we obtain the partition
function of infinitely diluted solution of proteins in water environment.
The calculated dependencies of the protein heat capacities upon temperature
are compared with the corresponding results of experimental measurements
for staphylococcal nuclease and metmyoglobin.
\end{abstract}

\maketitle

\section{Introduction}
Proteins are biological polymers consisting of elementary structural units, amino acids. Being synthesized
at ribosome, proteins are exposed to the cell interior where they fold into their unique three dimensional structure.
The process of forming the protein's three dimensional structure is called the process of protein folding.
The correct folding of protein is of crucial importance for the protein's proper functioning.
Despite numerous works devoted to investigation of protein folding this process is still not
entirely understood. The current state-of-the-art in experimental and theoretical
studies of the protein folding process are described in recent reviews
and references therein~\cite{Munoz07,Dill08,Onuchic04,Shakhnovich06,Prabhu06}.

In this paper we develop a novel theoretical method for the description of the protein folding
process which is based on the statistical mechanics principles.
Considering the process of protein folding as a first order
phase transition in a finite system, we present a statistical mechanics model for treating the
folding$\leftrightarrow$unfolding phase transition in
single-domain proteins. The suggested method is based on the
theory developed for the helix$\leftrightarrow$coil transition in polypeptides
discussed in \cite{Yakubovich07_theoryEPJD,Yakubovich06a_EPN,Yakubovich07_resultsEPJD,Yakubovich06a,
Yakubovich06b,Yakubovich06c,Yakubovich08_fluctuations,ISolovyov06a,ISolovyov06b,ISolovyov06c,Yakubovich_IJQC,Yakubovich_ISACC09}.
A way to construct a parameter-free partition
function for a system experiencing $\alpha$-helix$\leftrightarrow$random coil phase transition {\it in vacuo} was studied in \cite{Yakubovich07_theoryEPJD}.
In \cite{Yakubovich07_resultsEPJD} we have calculated potential energy surfaces (PES)
of polyalanines of different lengths with respect to their twisting
degrees of freedom. This was done within the framework of classical
molecular mechanics. The calculated PES were then used to construct a para\-meter--free
partition function of a polypeptide and
to derive various thermodynamical characteristics of alanine
polypeptides as a function of temperature and polypeptide length.

In this paper we construct
the partition function of a protein {\it in vacuo}, which
is the further generalization of the formalism developed in \cite{Yakubovich06a},
accounting
for folded, unfolded and prefolded states of the protein. This way of the construction of the
partition function is consistent with nucleation-condensation scenario of protein folding,
which is a very common scenario for globular proteins~\cite{Noetling08} and implies that at
the early stage of protein folding the native-like hydrophobic nucleus of protein is formed, while at
the later stages of the protein folding process all the rest of amino acids also attain the native-like conformation.

For the correct description of the protein folding in water environment it is of primary importance to consider the interactions
between the protein and the solvent molecules. The hydrophobic interactions are known to be
the most important driving forces of protein folding \cite{Kumar02}.
In the present work we present a way how one can construct the partition function of the
protein accounting for the interactions with solvent, i.e. accounting for the hydrophobic effect.
The most prominent feature of our approach is that it is developed for concrete
systems in contrast to various generalized and toy-models of protein folding process.

We treat the hydrophobic interactions in the system using the statistical mechanics
formalism developed in \cite{Scheraga_water04} for the description of the thermodynamical
properties of the solvation process
of aliphatic and aromatic hydrocarbons in water.
However, accounting solely for hydrophobic interactions is not sufficient for the
proper description of the energetics of all conformational states of the protein and
one has to take electrostatic interactions into account. In the present work the
electrostatic interactions are treated within a similar framework as described in \cite{Bakk02}.

We have applied the developed statistical mechanics model of protein folding
for two globular proteins, namely staphylococcal nuclease and
metmyoglobin. These proteins have
simple two-stage-like folding kinetics and demonstrate two folding$\leftrightarrow$unfolding
transitions, refereed as heat and cold denaturation~\cite{Griko88,Privalov97}. The comparison of the results of the theoretical
model with that of the experimental measurements shows the applicability of the suggested
formalism for an accurate description of various
thermodynamical characteristics in the system, e.g. heat denaturation, cold denaturation,
increase of the reminiscent heat capacity of the unfolded protein, etc.

Our paper is organized as follows. In Sec.~\ref{theory} we present the formalism
for the construction of the partition function of the protein in water environment
and justify the assumptions made on the system's properties. In Section~\ref{RandD}
we discuss the results obtained with our model for the description of folding$\leftrightarrow$unfolding
transition in staphylococcal nuclease and metmyoglobin. In Section~\ref{conclusions} we summarize the paper
and suggest several ways for a further development of the theoretical formalism.

\section{Theoretical Methods}
\subsection{Partition function of a protein}
\label{theory}
To study thermodynamic properties of the system one needs to
investigate its potential energy surface with respect to all the degrees
of freedom.
For the description of macromolecular systems, such as proteins, efficient model approaches are necessary.

The most relevant degrees of freedom in the protein folding process are the twisting degrees of
freedom along its backbone chain
\cite{Yakubovich06a,Yakubovich06a_EPN,He98a,He98b,Yakubovich06b,Yakubovich06c,ISolovyov06b,ISolovyov06c,Yakubovich07_theoryEPJD}. These degrees of
freedom are defined for each amino acid of the protein except
for the boundary ones and are described by two dihedral angles
$\varphi_i$ and $\psi_i$ (for definition of $\varphi_i$ and $\psi_i$ see e.g.~\cite{Yakubovich06a,Yakubovich06a_EPN,Yakubovich06b,Yakubovich06c,ISolovyov06b,ISolovyov06c,Yakubovich07_theoryEPJD}).

The degrees of freedom of a protein can be classified as stiff and soft ones. We call the degrees
of freedom corresponding to the variation of bond lengths, angles
and improper dihedral angles as
stiff, while degrees of freedom corresponding to the angles
$\varphi_i$ and $\psi_i$ are soft degrees of
freedom~\cite{Yakubovich07_theoryEPJD}. The stiff degrees of freedom can be treated within the
harmonic approximation, because the energies needed for a noticeable
structural rearrangement with respect to these degrees of
freedom are about several eV, which is significantly larger than the
characteristic thermal energy of the system (kT),
being at room temperature equal to $0.026$ eV
\cite{ISolovyov06a,ISolovyov06b,ISolovyov06c,GROMOS,AMBER,CHARMM}.

A Hamiltonian of a protein is constructed as a
sum of the potential, kinetic and vibrational energy terms. Assuming the harmonic approximation
for the stiff degrees of freedom it is possible
to derive the following expression for the partition function of a protein {\it in vacuo} being in a
particular conformational state $j$ \cite{Yakubovich07_theoryEPJD}:

\noindent
\begin{eqnarray}
Z_{j}&=&A_{j}(kT)^{3N-3-\frac{l_s}{2}}
     \int_{\varphi\in\Gamma_j}... \int_{\psi\in\Gamma_j}
     e^{-\epsilon_{j}(\{\varphi,\psi\})/kT}{\rm d}\varphi_1... {\rm d}\varphi_n{\rm d}\psi_1... {\rm d}\psi_n,
\label{part_func_gen}
\end{eqnarray}

\noindent where $T$ is the temperature, $k$ is the Boltzmann constant, $N$ is the total number of atoms in the protein, $l_s$ is the number of soft degrees of freedom,
$A_{j}$ is defined as follows:
\begin{eqnarray}
A_{j}&=&\left[ \frac{V_{j}\cdot
M^{3/2}\cdot\sqrt{I_{j}^{(1)}I_{j}^{(2)}I_{j}^{(3)}}
\prod_{i=1}^{l_s}\sqrt{\mu_i^{s(j)}}}{(2\pi)^{\frac{l_s}{2}}\pi
\hbar^{3N}\prod_{i=1}^{3N-6-l_s}\omega_i^{(j)}}\right].
\label{eq:Aj}
\end{eqnarray}
\noindent $A_j$ is a factor which depends on
the mass of the protein $M$, its three main
momenta of inertia $I_j^{(1,2,3)}$, specific volume $V_j$, the frequencies of the
stiff normal vibrational modes $\omega_i^{(j)}$ and on the generalized masses $\mu_i^{s(j)}$
corresponding to the soft degrees of freedom~\cite{Yakubovich07_theoryEPJD}.
$\epsilon_i$ in Eq.~(\ref{part_func_gen}) describes the potential
energy of the system corresponding to the variation of soft degrees of freedom. Integration in Eq.~(\ref{part_func_gen})
is performed over a certain part of a phase space of the system (a subspace $\Gamma_j$) corresponding to the soft degrees of freedom $\varphi$ and $\psi$.
The form of the partition function
in Eq.~(\ref{part_func_gen}) allows one to avoid the multidimensional integration over
the whole coordinate space and to reduce the integration only to the relevant parts of the phase space. $\epsilon_{j}$ in
Eq.~(\ref{part_func_gen}) denotes the potential energy surface of the protein as a function of
twisting degrees of freedom in the vicinity of protein's conformational state $j$.
Note that in general the proper choice of all the relevant conformations of protein and the corresponding set of $\Gamma_j$ is not a trivial task.

One can expect that the factors $A_j$ in Eq.~(\ref{part_func_gen}) depend on
the chosen conformation of the protein. However, due to the fact
that the values of specific volumes, momenta of inertia and
frequencies of normal vibration modes of the system in
different conformations are expected to be close
\cite{Krimm80,Yakubovich06a}, the values of $A_j$ in all
conformations become nearly equal, at least in the zero order harmonic
approximation, i.e. $A_j\equiv A$. Another simplification of the integration in Eq.~(\ref{part_func_gen}) comes from the statistical independence
of amino acids. We assume that within each conformational state $j$ all amino acids can be treated
statistically independently, i.e. the particular conformational state of $i$-th amino acid characterized by angles
$\varphi_i\in \Gamma_j$ and $\psi_i\in \Gamma_j$ does not influence the potential energy surface of all other amino acids, and vice versa.
This assumption is well applicable for rigid conformational states of the protein such as native state.
For the native state of a protein all atoms of the molecule move in harmonic potential
in the vicinity of their equilibrium positions. However, in unfolded states of the protein the flexibility
of the backbone chain leads to significant variations of the distances between atoms,
and consequently to a significant variation of interactions between atoms.
Accurate accounting (both analytical and computational) for the interactions between distant atoms
in the unfolded state of a protein is extremely difficult
(see Ref.~\cite{Cubrovic07} for analytical treatment of interactions in unfolded states of a protein).
In this work we assume that all amino acids in unfolded state of a protein move
in the identical mean field created by all the amino acids and leave the corrections
to this approximation for further considerations.

With the above mentioned assumptions the partition function of a protein $Z_p$ (without any solvent) reads as:

\noindent
\begin{eqnarray}
Z_p&=&A\cdot
(kT)^{3N-3-\frac{l_s}{2}}\sum_{j=1}^{\xi}\prod_{i=1}^{a}\int_{-\pi}^{\pi }\int_{-\pi}^{
\pi}
\exp\left({-\frac{\epsilon_i^{(j)}(\varphi_i,\psi_i)}{kT}}\right){\rm
d}\varphi_i {\rm d}\psi_i, \label{Z_step2}
\end{eqnarray}

\noindent where the summation over $j$ includes all $\xi$ statistically relevant conformations of the protein, $a$
is the number of amino acids in the protein and $\epsilon_i^{(j)}$ is the potential energy
surface as a function of twisting degrees of freedom $\varphi_i$ and $\psi_i$ of the $i$-th amino acid
in the $j$-th conformational state of the protein.
The exact construction of $\epsilon_a^{(j)}(\varphi_i,\psi_i)$ for various
conformational states of a particular protein will be discussed below.
We consider the angles $\varphi$ and $\psi$ as the only two soft degrees
of freedom in each amino acid of the protein, and therefore the total number of soft degrees of freedom of the protein $l_s=2a$.

Partition function in Eq.~(\ref{Z_step2}) can be further simplified if one assumes (i) that
each amino acid in the protein can exist only in two conformations: the native state conformation
and the random coil conformation; (ii) the potential
energy surfaces for all the amino acids are identical.
This assumption is applicable for both the native and the random coil state.
It is not very accurate for the description of thermodynamical properties of single amino acids, but
is reasonable for the treatment of thermodynamical properties of the entire protein.
The judgment of the quality of this assumption could be made on the basis of comparison
of the results obtained with its use with experimental data. Such comparison
is performed in Sec.~\ref{RandD} of this work.

Amino acids in a protein being in its native state
vibrate in a steep harmonic potential. Here we assume that the potential energy
profile of an amino acid in the native conformation should not be very sensitive to the type
of amino acid and thus can be taken as, e.g., the potential energy surface for an alanine
amino acid in the $\alpha$-helix conformation ~\cite{Yakubovich07_resultsEPJD}. Using the same arguments
the potential energy profile for an amino
acid in unfolded protein state can be approximated by e.g. the potential of alanine in the
unfolded state of alanine polypeptide (see Ref.~\cite{Yakubovich07_resultsEPJD} for discussion and analysis of alanine's potential energy surfaces).
Indeed, for an unfolded state of a protein it is reasonable to expect that once neglecting the long-range interactions
all the differences in the potential energy surfaces of various amino acids arise from the
steric overlap of the amino acids's radicals.
This is clearly seen on alanine's potential energy surface at values of $\varphi>0^{\circ}$
presented in Ref.~\cite{Yakubovich07_resultsEPJD}.
But the part of the potential energy surface at $\varphi>0^{\circ}$ gives a minor contribution
to the entropy of amino acid at room temperature.
This fact allows one to neglect all the differences in potential energy surfaces for different amino acids in an unfolded protein,
at least in the zero order approximation.
This assumption should be especially justified for proteins with the rigid
helix-rich native structure.
The staphylococcal nuclease,  which we study here has definitely high $\alpha$-helix content.
Another argument which allows to justify our assumption for a wider family of proteins is the rigidity of the protein's native structure.
Below, we validate the
assumptions made by performing the comparison of the results of our theoretical model
with the experimental data for $\alpha/\beta$ rich protein metmyoglobin obtained in~\cite{Privalov97}.

For the description of the folding $\leftrightarrow$ unfolding transition
in small globular proteins obeying simple two-state-like folding kinetics we assume that the protein can exist
in one of three states: completely folded state, completely unfolded state and
partially folded state where some amino acids from the flexible regions with no prominent secondary structure
are in the unfolded state, while other amino acids are in the folded conformation.
With this assumption the partition function of the protein reads as:

\noindent
\begin{eqnarray}
Z_p&&=Z_{0}+\sum_{i=a-\kappa}^{a}\frac{\kappa !}{(i-(a-\kappa)) !(a-i)!}Z_i,
\label{Z_combinatorial}
\end{eqnarray}

\noindent where $Z_i$ is defined in Eq.~(\ref{part_func_gen}),
$Z_{0}$ is the partition function of the protein in completely unfolded state, $a$ is the total number
of amino acids in a protein
and $\kappa$ is the number of amino acids in flexible regions.
The factorial term in Eq.~(\ref{Z_combinatorial}) accounts for the states in which various amino
acids from flexible regions independently attain the native conformation.
The summation in Eq.~(\ref{Z_combinatorial}) is performed over all partially folded
states of the protein, where
$a-\kappa$ is the minimal possible number of amino acids being in the folded state.
The factorial term describes the
number of ways to select $i-(a-\kappa)$ amino acids from
the flexible region of the protein consisting of $\kappa$ amino acids attaining native-like conformation.

Finally, the partition function of the protein {\it in vacuo}
has the following form:

\noindent
\begin{eqnarray}
Z_p&=&\tilde{Z}_p\cdot A(kT)^{3N-3-a},
\label{Z_vac}
\end{eqnarray}

\noindent where

\noindent
\begin{eqnarray}
\label{eq:Z_vac_E0}
\tilde{Z}_p&=&
Z_{u}^{a}+\sum_{i=a-\kappa}^{a}\frac{\kappa ! Z_b^i Z_{u}^{a-i}\exp{\left(i\cdot E_0/kT\right)}}{(i-(a-\kappa)) !(a-i)!}\\
\label{eq:Z_u}
Z_b&=&\int_{-\pi}^{\pi}\int_{-\pi}^{\pi} \exp{\left(- \frac{\epsilon_b(\varphi, \psi)}{k T}\right)}{\rm d}\varphi{\rm d}\psi\\
\label{eq:Z_b}
Z_{u}&=&\int_{-\pi}^{\pi}\int_{-\pi}^{\pi}\exp{\left(- \frac{\epsilon_{u}(\varphi,\psi)}{k T}\right)}{\rm d}\varphi{\rm d}\psi.
\end{eqnarray}

\noindent Here we omitted the trivial factor describing the motion of the protein center of mass,
which is of no significance for the problem considered,
$\epsilon_b(\varphi,\psi)$  (b stands for {\it{bound}}) is the potential
energy surface of an amino acid in the native conformation
and $\epsilon_{u}(\varphi,\psi)$ (u stands for {\it{unbound}}) is
the potential energy surface of an amino acid in the random coil conformation.
The potential energy profile of an
amino acid is calculated as a function of its twisting degrees of
freedom $\varphi$ and $\psi$.
Let us denote by $\epsilon_b^{0}$ and $\epsilon_u^{0}$ the
global minima on the potential energy surfaces of an amino acid in folded
and in unfolded conformations respectively. The potential energy of
an amino acid then reads as
$\epsilon^{0}_{u,b}+\epsilon_{u,b}(\varphi,\psi)$. $E_0$ in Eq.~(\ref{eq:Z_vac_E0}) is defined as
the energy difference between the global energy minima of the amino acid potential energy surfaces
corresponding to the folded and unfolded conformations, i.e. $E_0=\epsilon_u^{0} - \epsilon_b^{0}$.
The potential energy surfaces for amino acids as functions of angles $\varphi$ and $\psi$
were calculated and thoroughly analyzed in~\cite{Yakubovich07_resultsEPJD}.

In nature proteins perform their function in the aqueous environment.
Therefore the correct theoretical
description of the folding$\leftrightarrow$unfolding transition in water environment should account for solvent effects.

\subsection{Partition function of a protein in water environment}

In this section we evaluate $E_0$ and construct
the partition function for the protein in water environment.

The partition function of the infinitely diluted solution of proteins ${Z}$
can be constructed as follows:

\noindent
\begin{eqnarray}
{Z}=\sum_{j=1}^{\xi}\tilde{Z}_p^{(j)} Z_{W}^{(j)},
\label{Z_water1}
\end{eqnarray}

\noindent where
$Z_{W}^{(j)}$ is the partition function
of all water molecules in the $j$-th conformational state of a protein and $\tilde{Z}_p^{(j)}$
is the partition function of the protein in its $j$-th conformational state, in which
we further omit the factor describing the contribution of stiff degrees of freedom in the system.
This is done in order to simplify the expressions, because stiff degrees of freedom provide
a constant contribution to the heat capacity of the system since the heat capacity of the ensemble of
harmonic oscillators is constant. Below for the simplicity of notations we put $\tilde{Z}_p\equiv Z_p$.

There are two types of water molecules in the system: (i) molecules in pure water and
(ii) molecules interacting with the protein.
We assume that only the water molecules being in the vicinity of the protein's surface
are involved in the folding$\leftrightarrow$unfolding transition,
because they are affected by the variation of the hydrophobic surface of a protein.
This surface is equal to the protein's solvent accessible surface area (SASA) of the hydrophobic amino acids.
The number of interacting molecules is proportional to SASA and include only the molecules
from the first protein's solvation shell.
This area depends on the conformation of the protein.
The main contribution to the energy of the system caused by the variation of the protein's SASA
associated with the side-chain radicals of amino acids because the contribution to the free energy
assosiated
with solvation of protein's backbone is small~\cite{Ptizin_book}. Thus, in this work we pay
the main attention to the accounting for the SASA change arising due to the solvation of side chain radicals.

We treat all water molecules as statistically independent, i.e. the energy spectra of the states of a given molecule and its vibrational frequencies
do not depend on a particular state of all other water molecules.
Thus, the partition function of the whole system $Z$ can be factorized and reads as:
\noindent
\begin{eqnarray}
{Z}=\sum_{j=1}^{\xi}Z_p^{(j)} Z_{s}^{Y_c(j)}Z_{w}^{N_t-Y_c(j)},
\label{Z_water2}
\end{eqnarray}

\noindent where $\xi$ is the total number of states of a protein,
$Z_s$ is the partition function of a water molecule affected by the interaction
with the protein and $Z_w$ is the partition function of a water molecule in pure water.
$Y_c(j)$ is the number of water molecules interacting with the protein in the $j$-th conformational state.
$N_t$ is the total number of water molecules in the system. To simplify the expressions
we do not account for water molecules that do not
interact with the protein in any of its conformational states, i.e. $N_t=\max_j\{Y_c(j)\}$.

To construct the partition function of water we follow the formalism developed
in~\cite{Scheraga_water04} and refer only to the most essential details of that work. The partition function of a water molecule in pure water reads as:

\begin{eqnarray}
Z_{w}=\sum_{l=0}^{4}\left[\xi_lf_l\exp(-E_l/kT)\right],
\label{ham_func_water_10}
\end{eqnarray}
where the summation is performed over 5 possible states of a water molecule
(the states in which water molecule has 4,3,2,1 or 0 hydrogen bonds with the neighboring molecules).
$E_l$ are the energies of these states and $\xi_l$ are the combinatorial factors being equal to
1,4,6,4,1 for $l=0,1,2,3,4$, respectively. They describe the number of choices
to form a given number of hydrogen bonds. $f_l$ in Eq.~(\ref{ham_func_water_10}) describes the contribution
due to the partition function arising to to the translation and libration oscillations of the molecule. In the harmonic
approximation $f_l$ are equal to:
\begin{eqnarray}
f_l=\left[1-\exp(-h\nu_l^{(T)}/kT)\right]^{-3}\left[1-\exp(-h\nu_l^{(L)}/kT)\right]^{-3},
\label{ham_func_water_11}
\end{eqnarray}
where $\nu_l^{(T)}$ and $\nu_l^{(L)}$ are translation and libration motions frequencies of a water molecule
in its $l$-th state, respectively. These frequencies are calculated in Ref.~\cite{Scheraga_water04} and are given in
Table~\ref{tab:freq}.
The contribution of the internal vibrations of water molecules is not included
in Eq.~(\ref{ham_func_water_10}) because the frequencies of these vibrations are
practically not influenced by the interactions with surrounding water molecules.

\begin{table}
\begin{tabular}{l               l c c c c}
  \hline
 Number of hydrogen bonds & 0 & 1 & 2 & 3 & 4\\
 \hline
 Energy level, $E_i$ (kcal/mol)  & 6.670 & 4.970 & 3.870 & 2.030 & 0\\
 Energy level, $E_i^s$ (kcal/mol) & 6.431 & 4.731 & 3.631 & 1.791 & -0.564\\
 Translational frequencies, $\nu_i^{(T)}$, cm$^{-1}$  \hspace{1cm} & 26 & 86 & 61 & 57 & 210\\
 Librational frequencies, $\nu_i^{(L)}$, cm$^{-1}$ & 197 & 374 & 500 & 750 & 750\\
 \hline
\end{tabular}
\caption{Parameters of the partition function of water according to~\cite{Scheraga_water04}}
\label{tab:freq}
\end{table}

The partition function of a water molecule from the protein's first solvation shell reads as:
\begin{eqnarray}
Z_{s}=\sum_{l=0}^{4}\left[\xi_lf_l\exp(-E_l^{s}/kT)\right],
\label{ham_func_water_9}
\end{eqnarray}
where $f_l$ are defined in Eq.~(\ref{ham_func_water_11}) and $E_l^{s}$
denotes the energy levels of a water molecule interacting with aliphatic hydrocarbons of protein's amino acids.
Values of energies $E_l^s$ are given it Table~\ref{tab:freq}. For simplicity we treat all side-chain radicals of a protein as aliphatic hydrocarbons because
most of the protein's hydrophobic amino acids consist of aliphatic-like hydrocarbons.
It is possible to account for various types of side chain radicals by using the
experimental results of the measurements of the solvation free energies of amino acid radicals
from Ref.~\cite{Privalov93} and associated works.
However,
this correction will imply the reparametrization of the theory presented in~\cite{Scheraga_water04} and will lead to the introduction
of $\sim 20\cdot5$ additional parameters. Here we do not perform such a task
since this kind of improvement of the theory
would smear out the understanding of the principal physical factors underlying the protein folding$\leftrightarrow$unfolding transition.

In our theoretical model we also
account for the electrostatic interaction of protein's charged groups with water.
The presence of electrostatic field around the protein leads to the reorientation of H$_2$O molecules in the vicinity of charged groups
due to the interaction of dipole moments of the molecules with the electrostatic field. The additional factor arising
in the partition function~(\ref{ham_func_water_10})
of the molecules reads as:

\begin{eqnarray}
Z_{E}=\left(\frac{1}{4\pi}\int{\exp\left(-\frac{E\cdot d \cos{\theta}}{kT}\right)\sin{\theta}{\rm d}\theta{\rm d}\varphi}\right)^{\alpha},
\label{Z_eleccorr}
\end{eqnarray}
where $E$ is the strength of the electrostatic field,
$d$ is the absolute value of the H$_2$O molecule dipole moment,
$\alpha$ is the ratio of the number of water molecules
that interact with the electrostatic field of the protein ($N_E$) to
the number of water molecules interacting with the surface of the amino acids
from the inner part of the protein while they are exposed to water when
the protein is being unfolded ($N_w$), i.e. $\alpha=N_E/N_w$.
Note that
the effects of electrostatic interaction turn out to be more pronounced in the folded state of the protein.
This happens because in the unfolded state of a protein opposite charges of
amino acid's radicals are in average closer in space due to the flexibility of the backbone chain, while in the folded state
the positions of the charges are fixed by the rigid structure of a protein.

Integrating Eq.~(\ref{Z_eleccorr}) allows to write the factor $Z_E$ for the partition function of a
single H$_2$O molecule in pure water in the form:
\begin{eqnarray}
Z_{E}=\left(\frac{kT \sinh{\left[\frac{E d}{kT}\right]}}{E d}\right)^{\alpha}.
\label{Z_elec}
\end{eqnarray}

This equation shows how the electrostatic field enters the partition function.
In general, $E$ depends on the position in space with respect to the protein.
However, here we neglect this dependence and instead we treat the parameter $E$
as an average, characteristic electrostatic field created by the protein.

Let us denote by $N_s$ the number of water molecules interacting with the proteins surface in its folded state
i.e. $N_t=N_s+N_w$; where $N_t$ is defined in Eq~(\ref{Z_water2}).
We assume that the number of water molecules
interacting with the protein ($Y_c$) is linearly dependent on the number
of amino acids being in the unfolded conformation, i.e. $Y_c=N_s + i N_w/a$, where $i$ is the number
of the amino acids in the unfolded conformation and $a$ is the total number of amino acids
in the protein. Thus, the partition function~(\ref{Z_water2}) with the accounting for the factor~(\ref{Z_elec})
reads as:

\noindent
\begin{eqnarray}
\label{eq:Z_prefinal}
{Z}=Z_{s}^{N_s}\cdot \sum_{j=1}^{\xi}\left(Z_b Z_{w}^{N_w/a} Z_{E}^{N_w/a}\exp{\left(i\cdot E_0/kT\right)}\right
)^{i(j)}\left(Z_u Z_{s}^{N_w/a}\right)^{a-i(j)},
\end{eqnarray}
\noindent where $i(j)$ denotes the number of the amino acids being in the folded conformation when
the protein is in the $j$-th conformational
state. Accounting for the statistical factors for amino acids being in the folded and unfolded
states, similarly to how it was done for the vacuum case (see Eq.~(\ref{eq:Z_vac_E0})), one derives
from Eq.~(\ref{eq:Z_prefinal}) the following final expression:

\begin{eqnarray}
\label{eq:PF_final}
Z&=&\left(Z_s\right)^{N_s}\times\\
\nonumber &\times &\left[Z_u^a Z_s^{N_w}+\sum_{i=a-\kappa}^{a}\frac{\kappa ! \exp{\left(i\cdot E_0/kT\right)}}{(i-(a-\kappa)) !(a-i)!}
\left(Z_b Z_w^{N_w/a}Z_E^{N_w/a}\right)^i (Z_{u}Z_s^{N_w/a})^{a-i}\right],
\end{eqnarray}

\noindent where the term in the square brackets accounts for all statistically
significant conformational states of the protein.

Having constructed the partition function of the system
we can evaluate with its use all thermodynamic characteristics of the system, such as e.g. entropy, free energy, heat capacity, etc.
The free energy ($F$) and heat the capacity ($c$) of the system can be calculated from the partition function as follows:

\begin{eqnarray}
\label{eq:F}
F(T)=-k T \ln{Z(T)},\\
\label{eq:c}
c(T)=-T\frac{\partial ^2F(T)}{\partial T^2}.
\end{eqnarray}

In this work we analyze the dependence of protein's heat capacity
on temperature and compare the predictions of our model with available experimental data.

\section{Results and Discussion}
\label{RandD}

In this section we calculate the dependencies of the heat capacity on temperature
for two globular proteins metmyoglobin and staphylococcal nuclease and compare the results with
experimental data from~\cite{Griko88, Privalov97}.

\begin{figure}[!htb]
\includegraphics[width=6.in,clip]{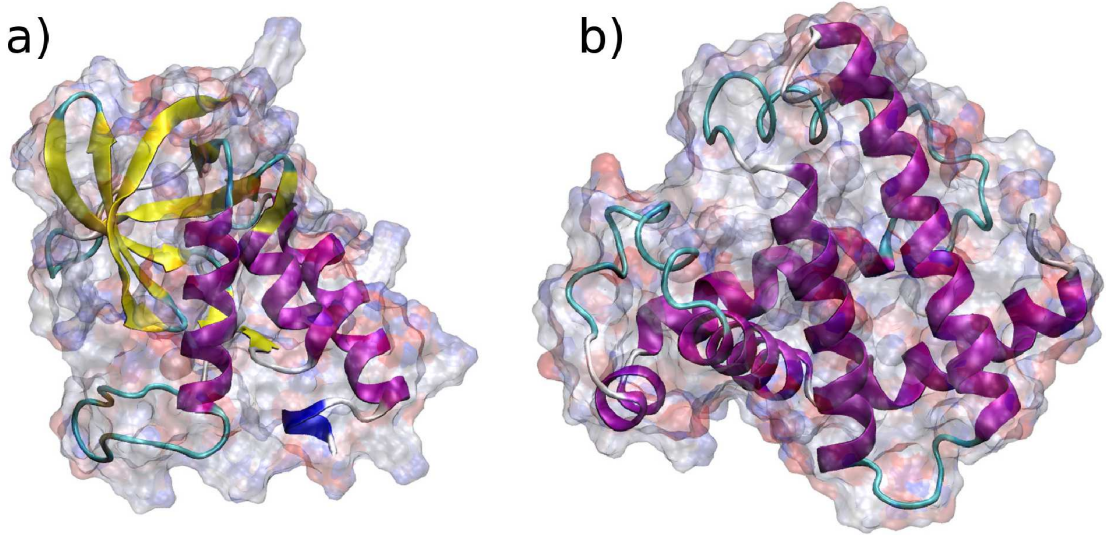}
\caption{a) Structure of staphylococcal nuclease (PDB ID 1EYD~\cite{1EYDPDB}),
and b) horse heart metmyoglobin (PDB ID 1YMB~\cite{1YMBPDB}). Images have been rendered using VMD program~\cite{VMD}.}
\label{fg:protsturct}
\end{figure}

The structures of metmyoglobin and staphylococcal nuclease proteins are shown in Fig.~\ref{fg:protsturct}. These are relatively
small globular proteins consisting of $\sim$150 amino acids. Under certain experimental conditions
(salt concentration and pH) the metmyoglobin and the staphylococcal nuclease experience two
folding$\leftrightarrow$unfolding transitions, which induce two peaks in the dependency of heat capacity on temperature
(see further discussion).
The peaks at lower temperature are due to the cold denaturation of the proteins. The peaks at higher
temperatures arise due to the ordinary folding$\leftrightarrow$unfolding transition.
The availability of
experimental data for the heat capacity profiles of the mentioned proteins, the presence of the cold denaturation and simple two-stage-like
folding kinetics are the reasons for selecting these particular proteins as case studies for the verification of the developed theoretical model.

\subsection{Heat capacity of staphylococcal nuclease}
Staphylococcal or micrococcal nuclease (S7 Nuclease) is a relatively nonspecific enzyme that digests single-stranded and double-stranded nucleic acids, but is more active on single-stranded substrates~\cite{SNase79}. This protein consists of 149 amino acids.
It's structure
is shown in Fig.~\ref{fg:protsturct}a.

To calculate the SASA of staphylococcal nuclease in the folded state the 3D structure of the
protein was taken from the Protein Data Bank~\cite{PDB} (PDB ID 1EYD). Using CHARMM27~\cite{CHARMM}
forcefield and NAMD program~\cite{NAMD} we performed the structural optimization of the protein
and calculated SASA with the solvent probe radius 1.4 \AA.

The value of SASA  of the side-chain radicals in the folded protein conformation
is equal to $S_f=$6858 \AA$^2$. In order to calculate SASA for an unfolded protein state,
the value of all angles $\varphi$ and $\psi$ were put equal to $180^{\circ}$,
corresponding to a fully stretched conformation.
Then, the optimization of the structure with the fixed angles $\varphi$ and $\psi$ was performed.
The optimized geometry of the stretched molecule has a minor dependence on the
value of dielectric susceptibility of the solvent, therefore the value of dielectric susceptibility was chosen to be equal to 20,
in order to mimic the screening of charges by the solvent. SASA of
the side-chain radicals in the stretched conformation of the protein is equal to $S_u=$15813 \AA$^2$.

The change of the number of water molecules those interacting with the protein due to the unfolding
process can be calculated as follows:

\begin{eqnarray}
N_w=(S_{u}-S_{f})n^{2/3},
\label{N_water}
\end{eqnarray}
where $S_u=15813$~\AA$^2$~and $S_f=6858$~\AA$^2$~are the SASA of the protein in unfolded and in folded conformations, respectively and
$n$ is the density of the water molecules.
The volume of one mole of water is equal to 18~cm$^{3}$, therefore $n\approx30$~\AA$^{-3}$

To account for the effects caused by the electrostatic interaction of water molecules with the charged groups
of the protein it is necessary to evaluate the strength
of the average electrostatic field $E$ in Eq.~(\ref{Z_elec}). The strength of the average field can be
estimated as $E\cdot d=kT$, where $d$ is the dipole moment
of a water molecule, $k$ is Bolzmann constant and T=300 K is the room temperature.
According to this estimate the energy of characteristic electrostatic interaction of water
molecules is equal to the thermal energy per degree of freedom of a molecule.

The total number of water molecules that interact with the electrostatic field of the protein can be
estimated from the known Debye screening length of a charge in electrolyte as follows:

\begin{eqnarray}
N_E=N_q\frac{4\pi \rho}{3} \lambda_d^3,
\label{eq:Debye}
\end{eqnarray}
\noindent where $N_q$ is the number of charged groups in the protein,
$\rho$ is the density of water and $\lambda$ is the Debye screening length.
Debye screening length of the symmetric electrolyte can be calculated as follows~\cite{Russel}:

\begin{eqnarray}
\lambda_d=\sqrt{\frac{\epsilon \epsilon_0 k T}{2 N_A e^2 I}},
\label{eq:Debye_length}
\end{eqnarray}

\noindent where $\epsilon_0$ is the permittivity of free space, $\epsilon$ is the dielectric constant,
$N_A$ is the Avogadro number, $e$ is the elementary charge and $I$ is
is the ionic strength of the electrolyte.

The experiments on denaturation of staphylococcal nuclease and metmyoglobin were performed in 100~mM ion buffer
of sodium chloride and 10mM buffer of sodium acetate respectively~\cite{Griko88,Privalov97}.
The Debye screening length in water with 10~mM and 100~mM concentration of
ions is $\lambda_d=$30~\AA~and $\lambda_d=$10~\AA~at room temperature respectively.

The described method allows to estimate the number of water molecules ($N_E$) interacting with
electric filed created by the charged groups of a protein. It should be considered as qualitative estimate
since we
have assumed the average electric field as being constant within a sphere of the radius $\lambda_d$, but
in fact it experiences some variations.
Thus, at the distances $\sim$15~\AA~from the point charge the interaction energy
of a H$_2$O molecule with the electric field becomes equal to $\sim0.02$~kT
(for this estimate we have used the linear growing distance-dependent
dielectric susceptibility $\epsilon=6 R$ as derived in~\cite{Mallik02} for the atoms fully exposed to the solvent).
However, we expect that the more accurate analysis accounting for the spatial variation
of the electric field will not change significantly the results of the analysis reported
here, because it is based on the physically correct picture of the effect
and the realistic values of all the physical quantities.
At physiological conditions staphylococcal nuclease has 8 charged residues~\cite{Zhou02}.
The value of $\alpha$ for this protein varies within the interval from 1.29 to 31.27 for
$\lambda_d \in$[10..30]~\AA. In our numerical analysis we have used the characteristic value of $\alpha$
equal to 2.5.

Note that number of molecules interacting with the electrostatic field $N_E$ and the strength of the electrostatic field $E$
should be considered as the $effective$ parameters of our model.
In this work we do not perform accurate accounting for the spatial dependence of the electrostatic field.
Instead, we introduce the parameters $\alpha$ and $E$ that can be interpreted as effective values
of the number of H$_2$O molecules and the strength of the electrostatic field correspondingly.
Let us stress that the number of water molecules $\alpha$ and the strength of the field $E$
are not independent parameters of our model because by choosing the higher value of $E$ and
smaller value of $\alpha$ or vice versa one can derive the same heat capacity profile.

In this work we do not investigate the dependencies of the heat capacity
profiles on the values of the parameters $\alpha$ and $E$. Below we
focus on the investigation of the dependence of the protein heat capacity on the energy $E_0$ at the fixed value of
$\alpha$ and $E$ equal to 2.5 and 0.58 kcal/mol respectively.

An important parameter of the model is the energy difference between the two states
of the protein normalized per one amino acid, $E_0$ introduced in Eq.~(\ref{eq:Z_vac_E0}).
This parameter describes both the energy loss due to the separation
of the hydrophobic groups of the protein which attract in the native state of the protein
due to Van-der-Waals interaction and
the energy gain due to the formation of Van-der-Waals interactions of
hydrophobic groups of the protein with H$_2$O molecules in the protein's unfolded state.
Also, the difference of the electrostatic energy of the system in the folded and unfolded states is accounted for in $E_0$.
The difference of the electrostatic energy may depend on various characteristics of the system, such as
concentration of ions in the solvent and its pH, on the exact location
of the charged sites in the native conformation of the protein and on the probability distribution
of distances between charged amino acids in the unfolded state.
Thus, exact calculation of $E_0$ is rather difficult. It is a separate task which we do not intend
to address in this work.
Instead, in the current study the energy difference
between the two phases of the protein is considered as a parameter of the model.
We treat $E_0$ as being dependent on external properties of the system, in particular on the pH value of the solution.

Another characteristic of the protein folding$\leftrightarrow$unfolding transition is its cooperativity.
In the model it is described by the parameter $\kappa$ in Eq.~(\ref{Z_combinatorial}). $\kappa$ describes
the number of amino acids in the flexible regions of the protein.
The staphylococcal nuclease possesses a prominent two-stage folding
kinetics, therefore only 5-10\% of amino acids is in the protein's flexible regions.
Thus, the value of $\kappa$ for this protein is small. It can be estimated as being equal to $149\cdot7\%\approx10$ amino acids.

The values of $E_0$
for staphylococcal nuclease at different values of pH are given in Table~\ref{tab:PHSN}.

\begin{table}
\begin{tabular}{l  r  r  r  r  r}
\hline
pH value & 7.0 & 5.0 & 4.5 & 3.88 & 3.23\\ [1ex]
\hline
$E_0$ (kcal/mol) & 0.789 & 0.795 & 0.803 & 0.819 & 0.890\\ [1ex]
\hline
\end{tabular}
\caption{Values of $E_0$ for staphylococcal nuclease at different values of pH of the solvent}
\label{tab:PHSN}
\end{table}

For the analysis of the variation
of the thermodynamic properties of the system during
the folding process one can omit all the contributions
to the free energy of the system that do not alter
significantly in the temperature range between -50$^ \circ$C and 150$^ \circ$.
Therefore, from the expression for the total free energy of the system $F$ we can subtract
all slowly varying contributions $F_0$ as follows:
\begin{eqnarray}
\label{eq:delta_F_1}
\delta F&=&F - F_0 = - (kT \ln{Z} - kT \ln {Z_0}) = - kT \ln{\left(\frac{Z}{Z_0}\right)},\\
\end{eqnarray}
From Eq.~(\ref{eq:delta_F_1}) follows that the subtraction of $F_0$
corresponds to the division of the total partition function $Z$ by the partition
function of the subsystem ($Z_0$) with slowly varying thermodynamical properties.
Therefore, in order to simplify the expressions, one can divide the partition
function in Eq.~(\ref{eq:PF_final}) by the partition function of
fully unfolded conformation of a protein (by $Z_u^a Z_s^{N_w}$) and by the partition
function of $N_s$ free water molecules (by $Z_w^{N_s}$). Thus, Eq.~(\ref{eq:PF_final}) can be
rewritten as follows:

\begin{eqnarray}
Z=\left(\frac{Z_s}{Z_w}\right)^{N_s}\left(1+\sum_{i=a-\kappa}^{a}
\frac{\kappa ! \exp\left({i\cdot E_0/kT}\right) }{(i-(a-\kappa)) !(a-i)!}\left(\frac{Z_b}{Z_u}\right)^i
\left(\frac{Z_wZ_E}{Z_s}\right)^{i N_w/a}\right).
\label{eq:PF_final_2}
\end{eqnarray}

With the use of Eq.~(\ref{eq:c}) on can calculate the heat capacity of the system as follows:

\begin{eqnarray}
c(T)=A+B(T-T_0)-T\frac{\partial ^2F(T)}{\partial T^2},
\label{eq:HC_AB}
\end{eqnarray}
\noindent where the factors $A$ and $B$ are responsible for the absolute value and the inclination
of the heat capacity curve respectively. These factors account for the contribution of stiff harmonic vibrational
modes in the system (factor $A$) and for the unharmonic correction to these vibrations
(factors $B$ and $T_0$). The contribution of protein's stiff vibrational
modes and the heat capacity of the fully unfolded conformation of protein
is also included into these factors. In our numerical analysis we have adjusted the values of $A$, $B$ and $T_0$
in order to match experimental measurements. However, factors $A$, $B$ and $T_0$
should not be considered as parameters of our model
since their values are not related to the thermodynamic
characteristics of the folding$\leftrightarrow$unfolding transition
and depend not entirely on the properties of the protein
but also on the properties of the solution, protein and ion concentrations, etc.

In our calculations for staphylococcal nuclease we have used the values of $A=1.25$~JK$^{-1}$g$^{-1}$, $B=6.25\cdot10^{-3}$~JK$^{-2}$g$^{-1}$ and $T_0=$323 $^\circ K$ in Eq.~(\ref{eq:HC_AB}).

\begin{figure}[!htb]
\begin{center}
\includegraphics[width=16cm,clip]{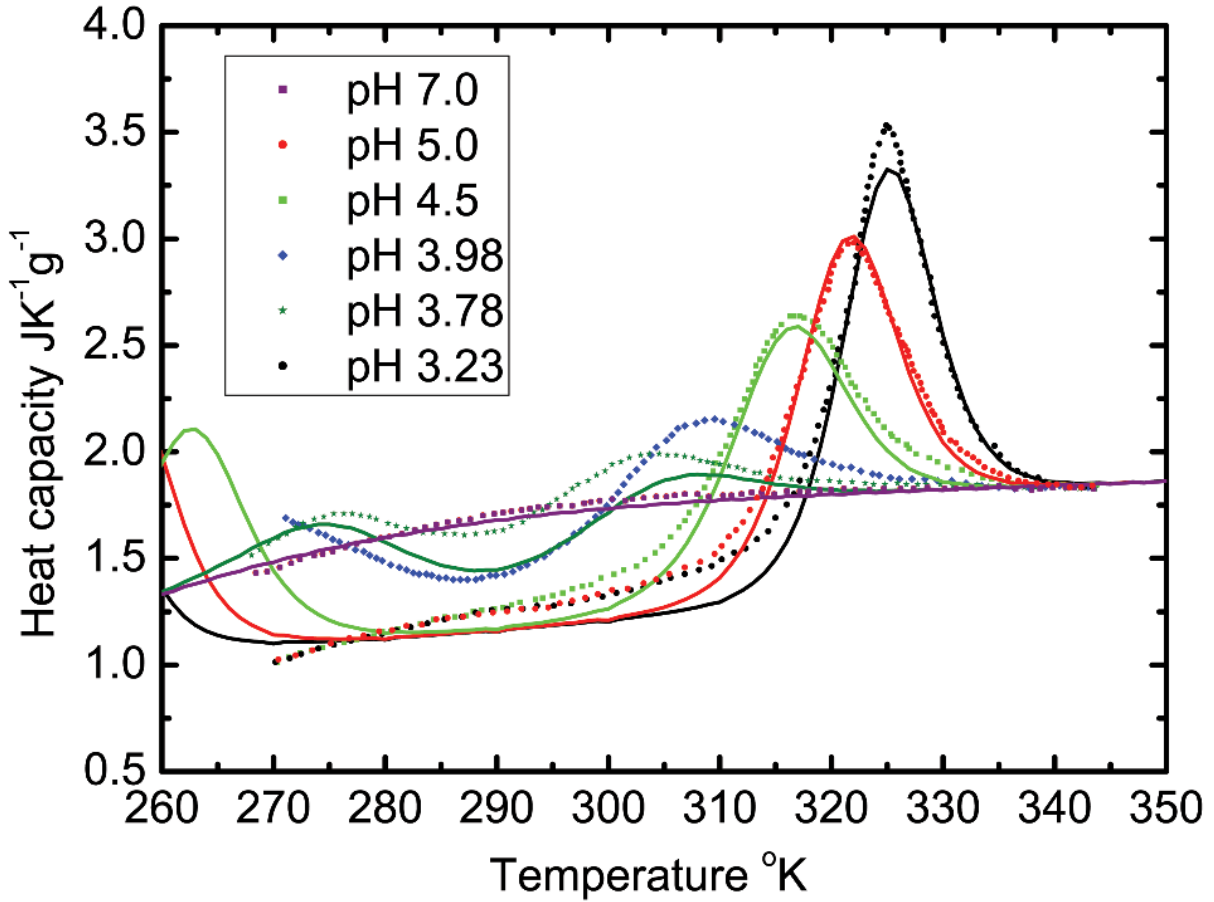}
\caption{Dependencies of the heat capacity on temperature for
staphylococcal nuclease (see Fig.~\ref{fg:protsturct}a) at different
values of pH.
Solid lines show results of the calculation, while
symbols present experimental data from Ref.~\cite{Griko88}.}
\end{center}
\label{fg:HC1EYD}
\end{figure}

The dependence of heat capacity on temperature
calculated for staphylococcal nuclease at different pH values are presented in
Fig.~\ref{fg:HC1EYD} by solid lines.
The results of experimental
measurements form Ref.~\cite{Griko88} are presented by symbols.
From Fig.~\ref{fg:HC1EYD} it is seen that staphylococcal nuclease experience
two folding$\leftrightarrow$transitions in the range of pH between 3.78 and 7.0.
At the pH value 3.23 no peaks in the heat capacity is present. It
means that the protein exists in the unfolded state over the whole
range of experimentally accessible temperatures.

Comparison of the theoretical results with experimental data shows
that our theoretical model reproduces experimental behavior better for the solvents with higher
pH. The heat capacity peak arising at higher temperatures due to the standard
folding$\leftrightarrow$unfolding transition is reproduced very well
for pH values being in the region 4.5-7.0. The deviations
at low temperatures can be attributed to the inaccuracy of the
statistical mechanics model of water in the vicinity of
the freezing point.

The accuracy of the statistical mechanics model for low pH values around 3.88
is also quite reasonable.
The deviation of theoretical curves from experimental ones likely
arise due to the alteration of the solvent properties at high concentration of protons
or due to the change of partial charge of amino acids at pH values being far
from the physiological conditions.

Despite some difference between the predictions of the developed model and
the experimental results arising
at certain temperatures and values of pH the overall performance of the model
can be considered as extremely good for such a complex process as structural folding transition
of a large biological molecule.

\subsection{Heat capacity of metmyoglobin}
Metmyoglobin is an oxidized form of a protein myoglobin. This is a monomeric protein containing
a single five-coordinate heme whose function is to reversibly form a dioxygen adduct~\cite{Myo04}. Metmyolobin
consists of 153 amino acids and it's structure is shown in Fig.~\ref{fg:protsturct} on the right.

In order to calculate SASA of side chain radicals of metmyoglobin
exactly the same procedure as for staphylococcal nuclease was performed (see discussion in the previous subsection).
SASA in the folded and unfolded states of the protein has been calculated and is equal 6847 \AA$^2$ and 16926 \AA$^2$ respectively.
Thus, there are $984$
H$_2$O molecules interacting with protein's hydrophobic surface in its unfolded state.

The electrostatic interaction of water molecules with metmyoglobin was accounted for in the same way
as for staphylococcal nuclease. The parameter $\alpha$ in Eq.~(\ref{Z_elec}) was chosen to be equal to 2.5.
With this we derive that 10950 H$_2$O molecules involve in the interaction
with the electrostatic field of metmyoglobin in its folded state.
The strength of the field was chosen the same as for staphylococcal nuclease.

The parameter $\kappa$ for metmyoglobin in Eq.~(\ref{Z_combinatorial}), describing the cooperativity of the
folding$\leftrightarrow$unfolding transition, differs significantly from
that for staphylococcal nuclease. The transition
in metmyoglobin is less cooperative than the transition in staphylococcal nuclease
because metmyoglobin has intermediate
partially folded states~\cite{Schortle01}.
Thus, while the rigid native-like core of the protein is formed,
a significant fraction of amino acids in the flexible regions of the protein
can exist in the unfolded state.
We assume  that ~1/3 of metmyoglobin's amino acids are in the flexible region,
i.e. the parameter $\kappa$ in Eq.~(\ref{Z_combinatorial})
equal to 50.

\begin{table}
\begin{tabular}{l  r  r  r  r}
\hline
pH value & 4.10 & 3.70 & 3.84 & 3.5 \\
\hline
$E_0$ (kcal/mol) & 1.128 & 1.150 & 1.165 & 1.2 \\
\hline
\end{tabular}
\caption{Values of $E_0$ for metmyoglobin at different values of solvent pH.}
\label{tab:MET}
\end{table}

The values of $E_0$ in Eq.~(\ref{eq:Z_vac_E0})
differ from that for staphylococcal nuclease and are compiled in Table~\ref{tab:MET}.
In our calculations for metmyoglobin we have used the values of $A=1.6$~JK$^{-1}$g$^{-1}$, $B=8.25\cdot10^{-3}$~JK$^{-2}$g$^{-1}$ and $T_0$=323 $^\circ K$ in Eq.~(\ref{eq:HC_AB}).

\begin{figure}[!htb]
\includegraphics[width=1.0\textwidth]{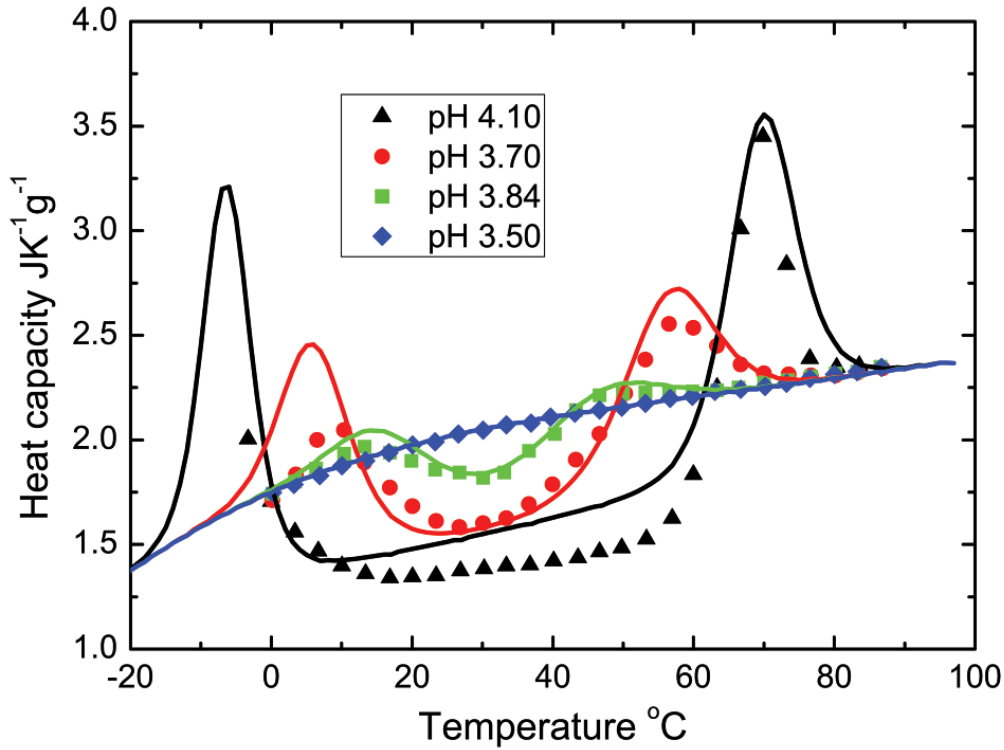}
\caption{Dependencies of the heat capacity on temperature for horse heart metmyoglobin
(see Fig.~\ref{fg:protsturct}b) at different
values of pH. Solid lines show the results of the calculation.
Symbols present the experimental data from
Ref.~\cite{Privalov97}.
}
\label{fg:HC1YMB}
\end{figure}

Solid lines in Fig.~\ref{fg:HC1YMB} show the dependence of the metmyoglobin's heat capacity on temperature
calculated using the developed theoretical model.
The experimental data from Ref.~\cite{Privalov97} are shown by symbols.

Metmyoglobin experiences two folding$\leftrightarrow$unfolding
transitions at the pH values exceeding 3.5 which can be called as cold and heat denaturations of the protein.
The dependence of the heat capacity on temperature therefore has two characteristic peaks, as seen in Fig.~\ref{fg:HC1YMB}.
Figure~\ref{fg:HC1YMB} shows that at pH lower than
3.84 metmyoglobin exists only in the unfolded state.

The comparison of predictions of the developed theoretical model with the
experimental data on heat capacity shows that
the theoretical model is well applicable for metmyoglobin case as well. The good agreement of
the theoretical and experimental heat capacity profiles over the whole range of temperatures and pH values
shows that the model treats correctly the thermodynamics of the protein folding process.

Our theory includes a number of parameters, namely
the energy difference between two phases $E_0$, strength of the electrostatic field $E$, number
of interacting H$_2$O molecules $\alpha$, the parameter describing the cooperativity of the phase
transition $\kappa$, as well as other parameters introduced in Ref.~\cite{Scheraga_water04} to treat the partition
function of water.
Three parameters, $E$, $E_0$ and $\kappa$, are
dependent on the properties of a particular protein and on the pH of the solvent. We have adjusted the values
of these parameters in order to reproduce the experimental data. All other
parameters of the model describing the structure of energy levels of water molecules,
their vibrational and librational frequencies, etc.
are considered as fixed, being universal for all proteins.

In spite of the model features of our approach, we want to stress that the complex behavior
and the peculiarities in dependencies of the heat capacity on temperature
are all very well reproduced by the developed model with only a few parameters.
This was demonstrated for two proteins and we consider this result as a significant
achievement.
This fact supports our conclusion that the developed model can be used for the prediction
of new features of phase transitions in various biomolecular systems.
Indeed, from Figs.~\ref{fg:HC1EYD} and~\ref{fg:HC1YMB} one can extract
a lot of useful information on the heat capacity profiles: the concave bending
of the heat capacity profile for a completely unfolded protein, the temperature of the cold and heat
denaturation, the absolute values of the heat capacity at the phase transition temperature, the
broadening of heat capacity peaks. Another peculiarity which is well
reproduced by our statistical mechanics model is the decrease of the heat capacity
of the folded state of the protein in comparison with that for unfolded state and asymmetry
of the heat capacity peaks.

\section{Conclusions}
\label{conclusions}

We have developed a novel statistical mechanics model for the description
of folding$\leftrightarrow$unfolding processes in globular proteins
obeying simple two-stage-like folding kinetics.
The model is based on the construction
of the partition function of the system as a sum over all
statistically significant conformational states of a protein.
The partition function of each state is a product of partition function
of a protein in a given conformational state, partition function
of water molecules in pure water and a partition function of H$_2$O molecules
interacting with the protein.

The introduced model includes
a number of parameters responsible for
certain physical properties of the system. The parameters were obtained from available
experimental data and three of them (energy difference between two phases, cooperativity of the transition
and the average strength of the protein's electrostatic field)
were considered as being variable depending on a particular protein and pH of the solvent.

We have compared the predictions of the developed model
with the results of experimental measurements of the dependence of the heat capacity on temperature
for staphylococcal nuclease
and metmyoglobin. The experimental results were obtained at
various pH of solvent.
The suggested model is capable to
reproduce well within a single framework a large number of peculiarities of the heat capacity profile,
such as the temperatures of cold and heat
denaturations, the corresponding maximum values of the heat capacities, the temperature range
of the cold and heat denaturation transitions, the difference between heat capacities of the folded
and unfolded states of the protein.

The good agreement of the results of calculations obtained using the developed
formalism with the results of experimental measurements
demonstrates that it can be used for the analysis of thermodynamical properties
of many biomolecular systems. Further development of the model can be focused on its
advance and application for the description of the influence of mutations on protein stability,
analysis of assembly and stability of protein complexes,
protein crystallization process, etc.

\section{Acknowledgments}
We acknowledge support of this work by the NoE EXCELL. We are grateful to Dr. Ilia Solov'yov for the careful reading of the manuscript and helpful advice.


\begin{thebibliography}{10}%
\makeatletter
\providecommand \@ifxundefined [1]{%
 \ifx #1\undefined \expandafter \@firstoftwo
 \else \expandafter \@secondoftwo
\fi
}%
\providecommand \@ifnum [1]{%
 \ifnum #1\expandafter \@firstoftwo
 \else \expandafter \@secondoftwo
\fi
}%
\providecommand \enquote [1]{``#1''}%
\providecommand \bibnamefont  [1]{#1}%
\providecommand \bibfnamefont [1]{#1}%
\providecommand \citenamefont [1]{#1}%
\providecommand\href[0]{\@sanitize\@href}%
\providecommand\@href[1]{\endgroup\@@startlink{#1}\endgroup\@@href}%
\providecommand\@@href[1]{#1\@@endlink}%
\providecommand \@sanitize [0]{\begingroup\catcode`\&12\catcode`\#12\relax}%
\@ifxundefined \pdfoutput {\@firstoftwo}{%
 \@ifnum{\z@=\pdfoutput}{\@firstoftwo}{\@secondoftwo}%
}{%
 \providecommand\@@startlink[1]{\leavevmode}%
 \providecommand\@@endlink[0]{}%
}{%
 \providecommand\@@startlink[1]{%
  \leavevmode
  \pdfstartlink
   attr{/Border[0 0 1 ]/H/I/C[0 1 1]}%
   user{/Subtype/Link/A<</Type/Action/S/URI/URI(#1)>>}%
  \relax
 }%
 \providecommand\@@endlink[0]{\pdfendlink}%
}%
\providecommand \url  [0]{\begingroup\@sanitize \@url }%
\providecommand \@url [1]{\endgroup\@href {#1}{\urlprefix}}%
\providecommand \urlprefix [0]{URL }%
\providecommand \Eprint[0]{\href }%
\@ifxundefined \urlstyle {%
  \providecommand \doi [1]{doi:\discretionary{}{}{}#1}%
}{%
  \providecommand \doi [0]{doi:\discretionary{}{}{}\begingroup
  \urlstyle{rm}\Url }%
}%
\providecommand \doibase [0]{http://dx.doi.org/}%
\providecommand \Doi[1]{\href{\doibase#1}}%
\providecommand \selectlanguage [0]{\@gobble}%
\providecommand \bibinfo [0]{\@secondoftwo}%
\providecommand \bibfield [0]{\@secondoftwo}%
\providecommand \translation [1]{[#1]}%
\providecommand \BibitemOpen[0]{}%
\providecommand \bibitemStop [0]{}%
\providecommand \bibitemNoStop [0]{.\EOS\space}%
\providecommand \EOS [0]{\spacefactor3000\relax}%
\providecommand \BibitemShut [1]{\csname bibitem#1\endcsname}%
\bibitem{Munoz07}%
  \BibitemOpen
  \bibfield{author}{%
  \bibinfo {author} {\bibfnamefont{V.}~\bibnamefont{Mu{\~n}oz}},\ }%
  \bibfield{journal}{%
  \bibinfo {journal} {Annu.\ Rev.\ Biophys.\ Biomol.\ Struct.}\ }%
  \textbf{\bibinfo {volume} {36}},\ \bibinfo {pages} {395} (\bibinfo {year}
  {2007})\BibitemShut{NoStop}%
\bibitem{Dill08}%
  \BibitemOpen
  \bibfield{author}{%
  \bibinfo {author} {\bibfnamefont{K.~A.}\ \bibnamefont{Dill}}, \bibinfo
  {author} {\bibfnamefont{S.~B.}\ \bibnamefont{Ozkan}}, \bibinfo {author}
  {\bibfnamefont{M.~S.}\ \bibnamefont{Shell}},\ and\ \bibinfo {author}
  {\bibfnamefont{T.~R.}\ \bibnamefont{Weikl}},\ }%
  \bibfield{journal}{%
  \bibinfo {journal} {Annu. Rev. Biophys.}\ }%
  \textbf{\bibinfo {volume} {37}},\ \bibinfo {pages} {289} (\bibinfo {year}
  {2008})\BibitemShut{NoStop}%
\bibitem{Onuchic04}%
  \BibitemOpen
  \bibfield{author}{%
  \bibinfo {author} {\bibfnamefont{J.~N.}\ \bibnamefont{Onuchic}}\ and\
  \bibinfo {author} {\bibfnamefont{P.~G.}\ \bibnamefont{Wolynes}},\ }%
  \bibfield{journal}{%
  \bibinfo {journal} {Curr. Op. Struct. Biol.}\ }%
  \textbf{\bibinfo {volume} {14}},\ \bibinfo {pages} {70} (\bibinfo {year}
  {2004})\BibitemShut{NoStop}%
\bibitem{Shakhnovich06}%
  \BibitemOpen
  \bibfield{author}{%
  \bibinfo {author} {\bibfnamefont{E.}~\bibnamefont{Shakhnovich}},\ }%
  \bibfield{journal}{%
  \bibinfo {journal} {Chem.\ Rev.}\ }%
  \textbf{\bibinfo {volume} {106}},\ \bibinfo {pages} {1559} (\bibinfo {year}
  {2006})\BibitemShut{NoStop}%
\bibitem{Prabhu06}%
  \BibitemOpen
  \bibfield{author}{%
  \bibinfo {author} {\bibfnamefont{N.~V.}\ \bibnamefont{Prabhu}}\ and\ \bibinfo
  {author} {\bibfnamefont{K.~A.}\ \bibnamefont{Sharp}},\ }%
  \bibfield{journal}{%
  \bibinfo {journal} {Chem. Rev}\ }%
  \textbf{\bibinfo {volume} {106}},\ \bibinfo {pages} {1616} (\bibinfo {year}
  {2006})\BibitemShut{NoStop}%
\bibitem{Yakubovich07_theoryEPJD}%
  \BibitemOpen
  \bibfield{author}{%
  \bibinfo {author} {\bibfnamefont{A.}~\bibnamefont{Yakubovich}}, \bibinfo
  {author} {\bibfnamefont{I.}~\bibnamefont{Solov'yov}}, \bibinfo {author}
  {\bibfnamefont{A.}~\bibnamefont{Solov'yov}},\ and\ \bibinfo {author}
  {\bibfnamefont{W.}~\bibnamefont{Greiner}},\ }%
  \bibfield{journal}{%
  \bibinfo {journal} {Eur.\ Phys.\ J.\ D}\ }%
  \textbf{\bibinfo {volume} {46}},\ \bibinfo {pages} {215} (\bibinfo {year}
  {2007}),\ \bibinfo {note} {arXiv:0704.3079v1 [physics.bio-ph], 23 Apr
  2007}\BibitemShut{NoStop}%
\bibitem{Yakubovich06a_EPN}%
  \BibitemOpen
  \bibfield{author}{%
  \bibinfo {author} {\bibfnamefont{A.}~\bibnamefont{Yakubovich}}, \bibinfo
  {author} {\bibfnamefont{I.}~\bibnamefont{Solov'yov}}, \bibinfo {author}
  {\bibfnamefont{A.}~\bibnamefont{Solov'yov}},\ and\ \bibinfo {author}
  {\bibfnamefont{W.}~\bibnamefont{Greiner}},\ }%
  \bibfield{journal}{%
  \bibinfo {journal} {Europhys.\ News}\ }%
  \textbf{\bibinfo {volume} {38}},\ \bibinfo {pages} {10} (\bibinfo {year}
  {2007})\BibitemShut{NoStop}%
\bibitem{Yakubovich07_resultsEPJD}%
  \BibitemOpen
  \bibfield{author}{%
  \bibinfo {author} {\bibfnamefont{I.}~\bibnamefont{Solov'yov}}, \bibinfo
  {author} {\bibfnamefont{A.}~\bibnamefont{Yakubovich}}, \bibinfo {author}
  {\bibfnamefont{A.}~\bibnamefont{Solov'yov}},\ and\ \bibinfo {author}
  {\bibfnamefont{W.}~\bibnamefont{Greiner}},\ }%
  \bibfield{journal}{%
  \bibinfo {journal} {Eur.\ Phys.\ J.\ D}\ }%
  \textbf{\bibinfo {volume} {46}},\ \bibinfo {pages} {227} (\bibinfo {year}
  {2008}),\ \bibinfo {note} {arXiv:0704.3085v1 [physics.bio-ph], 23 Apr
  2007}\BibitemShut{NoStop}%
\bibitem{Yakubovich06a}%
  \BibitemOpen
  \bibfield{author}{%
  \bibinfo {author} {\bibfnamefont{A.}~\bibnamefont{Yakubovich}}, \bibinfo
  {author} {\bibfnamefont{I.}~\bibnamefont{Solov'yov}}, \bibinfo {author}
  {\bibfnamefont{A.}~\bibnamefont{Solov'yov}},\ and\ \bibinfo {author}
  {\bibfnamefont{W.}~\bibnamefont{Greiner}},\ }%
  \bibfield{journal}{%
  \bibinfo {journal} {Eur.\ Phys.\ J.\ D}\ }%
  \textbf{\bibinfo {volume} {40}},\ \bibinfo {pages} {363} (\bibinfo {year}
  {2006})\BibitemShut{NoStop}%
\bibitem{Yakubovich06b}%
  \BibitemOpen
  \bibfield{author}{%
  \bibinfo {author} {\bibfnamefont{A.}~\bibnamefont{Yakubovich}}, \bibinfo
  {author} {\bibfnamefont{I.}~\bibnamefont{Solov'yov}}, \bibinfo {author}
  {\bibfnamefont{A.}~\bibnamefont{Solov'yov}},\ and\ \bibinfo {author}
  {\bibfnamefont{W.}~\bibnamefont{Greiner}},\ }%
  \bibfield{journal}{%
  \bibinfo {journal} {Eur.\ Phys.\ J.\ D}\ }%
  \textbf{\bibinfo {volume} {39}},\ \bibinfo {pages} {23} (\bibinfo {year}
  {2006})\BibitemShut{NoStop}%
\bibitem{Yakubovich06c}%
  \BibitemOpen
  \bibfield{author}{%
  \bibinfo {author} {\bibfnamefont{A.}~\bibnamefont{Yakubovich}}, \bibinfo
  {author} {\bibfnamefont{I.}~\bibnamefont{Solov'yov}}, \bibinfo {author}
  {\bibfnamefont{A.}~\bibnamefont{Solov'yov}},\ and\ \bibinfo {author}
  {\bibnamefont{W.Greiner}},\ }%
  \bibfield{journal}{%
  \bibinfo {journal} {Khimicheskaya Fizika (Chemical Physics) (in Russian)}\ }%
  \textbf{\bibinfo {volume} {25}},\ \bibinfo {pages} {11} (\bibinfo {year}
  {2006})\BibitemShut{NoStop}%
\bibitem{Yakubovich08_fluctuations}%
  \BibitemOpen
  \bibfield{author}{%
  \bibinfo {author} {\bibfnamefont{A.}~\bibnamefont{Yakubovich}}, \bibinfo
  {author} {\bibfnamefont{I.}~\bibnamefont{Solov'yov}}, \bibinfo {author}
  {\bibfnamefont{A.}~\bibnamefont{Solov'yov}},\ and\ \bibinfo {author}
  {\bibfnamefont{W.}~\bibnamefont{Greiner}},\ }%
  \bibfield{journal}{%
  \bibinfo {journal} {Eur.\ Phys.\ J.\ D}\ }%
  \textbf{\bibinfo {volume} {DOI: 10.1140/epjd/e2008-00126-y}} (\bibinfo {year}
  {2008})\BibitemShut{NoStop}%
\bibitem{ISolovyov06a}%
  \BibitemOpen
  \bibfield{author}{%
  \bibinfo {author} {\bibfnamefont{I.}~\bibnamefont{Solov'yov}}, \bibinfo
  {author} {\bibfnamefont{A.}~\bibnamefont{Yakubovich}}, \bibinfo {author}
  {\bibfnamefont{A.}~\bibnamefont{Solov'yov}},\ and\ \bibinfo {author}
  {\bibfnamefont{W.}~\bibnamefont{Greiner}},\ }%
  \bibfield{journal}{%
  \bibinfo {journal} {J.\ Exp.\ Theor.\ Phys.}\ }%
  \textbf{\bibinfo {volume} {103}},\ \bibinfo {pages} {463} (\bibinfo {year}
  {2006})\BibitemShut{NoStop}%
\bibitem{ISolovyov06b}%
  \BibitemOpen
  \bibfield{author}{%
  \bibinfo {author} {\bibfnamefont{I.}~\bibnamefont{Solov'yov}}, \bibinfo
  {author} {\bibfnamefont{A.}~\bibnamefont{Yakubovich}}, \bibinfo {author}
  {\bibfnamefont{A.}~\bibnamefont{Solov'yov}},\ and\ \bibinfo {author}
  {\bibfnamefont{W.}~\bibnamefont{Greiner}},\ }%
  \bibfield{journal}{%
  \bibinfo {journal} {Phys.\ Rev.~E}\ }%
  \textbf{\bibinfo {volume} {73}},\ \bibinfo {pages} {021916} (\bibinfo {year}
  {2006})\BibitemShut{NoStop}%
\bibitem{ISolovyov06c}%
  \BibitemOpen
  \bibfield{author}{%
  \bibinfo {author} {\bibfnamefont{I.}~\bibnamefont{Solov'yov}}, \bibinfo
  {author} {\bibfnamefont{A.}~\bibnamefont{Yakubovich}}, \bibinfo {author}
  {\bibfnamefont{A.}~\bibnamefont{Solov'yov}},\ and\ \bibinfo {author}
  {\bibfnamefont{W.}~\bibnamefont{Greiner}},\ }%
  \bibfield{journal}{%
  \bibinfo {journal} {J.\ Exp.\ Theor.\ Phys.}\ }%
  \textbf{\bibinfo {volume} {102}},\ \bibinfo {pages} {314} (\bibinfo {year}
  {2006})\BibitemShut{NoStop}%
\bibitem{Yakubovich_IJQC}%
  \BibitemOpen
  \bibfield{author}{%
  \bibinfo {author} {\bibfnamefont{A.}~\bibnamefont{Yakubovich}}, \bibinfo
  {author} {\bibfnamefont{A.}~\bibnamefont{Solov'yov}},\ and\ \bibinfo {author}
  {\bibfnamefont{W.}~\bibnamefont{Greiner}},\ }%
  \bibfield{journal}{%
  \bibinfo {journal} {Int. J. Quant. Chem.}\ }%
  \textbf{\bibinfo {volume} {110}},\ \bibinfo {pages} {257} (\bibinfo {year}
  {2010})\BibitemShut{NoStop}%
\bibitem{Yakubovich_ISACC09}%
  \BibitemOpen
  \bibfield{author}{%
  \bibinfo {author} {\bibfnamefont{A.}~\bibnamefont{Yakubovich}}, \bibinfo
  {author} {\bibfnamefont{A.}~\bibnamefont{Solov'yov}},\ and\ \bibinfo {author}
  {\bibfnamefont{W.}~\bibnamefont{Greiner}},\ }%
  \bibfield{journal}{%
  \bibinfo {journal} {AIP Conf. Proc.}\ }%
  \textbf{\bibinfo {volume} {1197}},\ \bibinfo {pages} {186} (\bibinfo {year}
  {2009})\BibitemShut{NoStop}%
\bibitem{Noetling08}%
  \BibitemOpen
  \bibfield{author}{%
  \bibinfo {author} {\bibfnamefont{B.}~\bibnamefont{Noetling}}\ and\ \bibinfo
  {author} {\bibfnamefont{D.~A.}\ \bibnamefont{Agard}},\ }%
  \bibfield{journal}{%
  \bibinfo {journal} {Proteins}\ }%
  \textbf{\bibinfo {volume} {73}},\ \bibinfo {pages} {754} (\bibinfo {year}
  {2008})\BibitemShut{NoStop}%
\bibitem{Kumar02}%
  \BibitemOpen
  \bibfield{author}{%
  \bibinfo {author} {\bibfnamefont{S.}~\bibnamefont{Kumar}}, \bibinfo {author}
  {\bibfnamefont{C.-J.}\ \bibnamefont{Tsai}},\ and\ \bibinfo {author}
  {\bibfnamefont{R.}~\bibnamefont{Nussinov}},\ }%
  \bibfield{journal}{%
  \bibinfo {journal} {Biochemistry}\ }%
  \textbf{\bibinfo {volume} {41}},\ \bibinfo {pages} {5359} (\bibinfo {year}
  {2002})\BibitemShut{NoStop}%
\bibitem{Scheraga_water04}%
  \BibitemOpen
  \bibfield{author}{%
  \bibinfo {author} {\bibfnamefont{J.~H.}\ \bibnamefont{Griffith}}\ and\
  \bibinfo {author} {\bibfnamefont{H.}~\bibnamefont{Scheraga}},\ }%
  \bibfield{journal}{%
  \bibinfo {journal} {J.~Mol.\ Struc.}\ }%
  \textbf{\bibinfo {volume} {682}},\ \bibinfo {pages} {97} (\bibinfo {year}
  {2004})\BibitemShut{NoStop}%
\bibitem{Bakk02}%
  \BibitemOpen
  \bibfield{author}{%
  \bibinfo {author} {\bibfnamefont{A.}~\bibnamefont{Bakk}}, \bibinfo {author}
  {\bibfnamefont{J.~S.}\ \bibnamefont{Hye}},\ and\ \bibinfo {author}
  {\bibfnamefont{A.}~\bibnamefont{Hansen}},\ }%
  \bibfield{journal}{%
  \bibinfo {journal} {BJ}\ }%
  \textbf{\bibinfo {volume} {82}},\ \bibinfo {pages} {713719} (\bibinfo {year}
  {2002})\BibitemShut{NoStop}%
\bibitem{Griko88}%
  \BibitemOpen
  \bibfield{author}{%
  \bibinfo {author} {\bibfnamefont{Y.}~\bibnamefont{Griko}}, \bibinfo {author}
  {\bibfnamefont{P.}~\bibnamefont{Privalov}}, \bibinfo {author}
  {\bibfnamefont{J.}~\bibnamefont{Aturtevant}},\ and\ \bibinfo {author}
  {\bibfnamefont{S.}~\bibnamefont{Venyaminov}},\ }%
  \bibfield{journal}{%
  \bibinfo {journal} {Proc.\ Natl.\ Acad.\ Sci.\ USA}\ }%
  \textbf{\bibinfo {volume} {85}},\ \bibinfo {pages} {3343} (\bibinfo {year}
  {1988})\BibitemShut{NoStop}%
\bibitem{Privalov97}%
  \BibitemOpen
  \bibfield{author}{%
  \bibinfo {author} {\bibfnamefont{P.}~\bibnamefont{Privalov}},\ }%
  \bibfield{journal}{%
  \bibinfo {journal} {J. Chem. Thermodyn.}\ }%
  \textbf{\bibinfo {volume} {29}},\ \bibinfo {pages} {447} (\bibinfo {year}
  {1997})\BibitemShut{NoStop}%
\bibitem{He98a}%
  \BibitemOpen
  \bibfield{author}{%
  \bibinfo {author} {\bibfnamefont{S.}~\bibnamefont{He}}\ and\ \bibinfo
  {author} {\bibfnamefont{H.~A.}\ \bibnamefont{Scheraga}},\ }%
  \bibfield{journal}{%
  \bibinfo {journal} {J.~Chem.\ Phys.}\ }%
  \textbf{\bibinfo {volume} {108}},\ \bibinfo {pages} {271} (\bibinfo {year}
  {1998})\BibitemShut{NoStop}%
\bibitem{He98b}%
  \BibitemOpen
  \bibfield{author}{%
  \bibinfo {author} {\bibfnamefont{S.}~\bibnamefont{He}}\ and\ \bibinfo
  {author} {\bibfnamefont{H.~A.}\ \bibnamefont{Scheraga}},\ }%
  \bibfield{journal}{%
  \bibinfo {journal} {J.~Chem.\ Phys.}\ }%
  \textbf{\bibinfo {volume} {108}},\ \bibinfo {pages} {287} (\bibinfo {year}
  {1998})\BibitemShut{NoStop}%
\bibitem{GROMOS}%
  \BibitemOpen
  \bibfield{author}{%
  \bibinfo {author} {\bibfnamefont{W.}~\bibnamefont{Scott}}\ and\ \bibinfo
  {author} {\bibfnamefont{W.}~\bibnamefont{van Gunsteren}},\ }%
  in\ \emph{\bibinfo {booktitle} {Methods and Techniques in Computational
  Chemistry: METECC-95}},\ \bibinfo {editor} {edited by\ \bibinfo {editor}
  {\bibfnamefont{E.}~\bibnamefont{Clementi}}\ and\ \bibinfo {editor}
  {\bibfnamefont{G.}~\bibnamefont{Corongiu}}}\ (\bibinfo {publisher} {STEF,
  Cagliari, Italy},\ \bibinfo {year} {1995})\ pp.\ \bibinfo {pages}
  {397--434}\BibitemShut{NoStop}%
\bibitem{AMBER}%
  \BibitemOpen
  \bibfield{author}{%
  \bibinfo {author} {\bibfnamefont{W.}~\bibnamefont{Cornell}}, \bibinfo
  {author} {\bibfnamefont{P.}~\bibnamefont{Cieplak}}, \bibinfo {author}
  {\bibfnamefont{C.}~\bibnamefont{Bayly}},\ and\ \bibinfo {author}
  {\bibnamefont{{\it et al}}},\ }%
  \bibfield{journal}{%
  \bibinfo {journal} {J.~Am.\ Chem.\ Soc.}\ }%
  \textbf{\bibinfo {volume} {117}},\ \bibinfo {pages} {5179} (\bibinfo {year}
  {1995})\BibitemShut{NoStop}%
\bibitem{CHARMM}%
  \BibitemOpen
  \bibfield{author}{%
  \bibinfo {author} {\bibfnamefont{A.}~\bibnamefont{MacKerell}}, \bibinfo
  {author} {\bibfnamefont{D.}~\bibnamefont{Bashford}}, \bibinfo {author}
  {\bibfnamefont{R.}~\bibnamefont{Bellott}},\ and\ \bibinfo {author}
  {\bibnamefont{{\it et al}}},\ }%
  \bibfield{journal}{%
  \bibinfo {journal} {J.~Phys.\ Chem.~B}\ }%
  \textbf{\bibinfo {volume} {102}},\ \bibinfo {pages} {3586} (\bibinfo {year}
  {1998})\BibitemShut{NoStop}%
\bibitem{Krimm80}%
  \BibitemOpen
  \bibfield{author}{%
  \bibinfo {author} {\bibfnamefont{S.}~\bibnamefont{Krimm}}\ and\ \bibinfo
  {author} {\bibfnamefont{J.}~\bibnamefont{Bandekar}},\ }%
  \bibfield{journal}{%
  \bibinfo {journal} {Biopolymers}\ }%
  \textbf{\bibinfo {volume} {19}},\ \bibinfo {pages} {1} (\bibinfo {year}
  {1980})\BibitemShut{NoStop}%
\bibitem{Cubrovic07}%
  \BibitemOpen
  \bibfield{author}{%
  \bibinfo {author} {\bibfnamefont{M.}~\bibnamefont{Cubrovic}}, \bibinfo
  {author} {\bibfnamefont{O.}~\bibnamefont{Obolensky}},\ and\ \bibinfo {author}
  {\bibfnamefont{A.}~\bibnamefont{Solov'yov}},\ }%
  \bibfield{journal}{%
  \bibinfo {journal} {Eur.\ Phys.\ J.\ D}\ }%
  \textbf{\bibinfo {volume} {51}},\ \bibinfo {pages} {41} (\bibinfo {year}
  {2009})\BibitemShut{NoStop}%
\bibitem{Ptizin_book}%
  \BibitemOpen
  \bibfield{author}{%
  \bibinfo {author} {\bibfnamefont{A.}~\bibnamefont{Finkelstein}}\ and\
  \bibinfo {author} {\bibfnamefont{O.}~\bibnamefont{Ptitsyn}},\ }%
  \emph{\bibinfo {title} {Protein Physics. A Course of Lectures}}\ (\bibinfo
  {publisher} {Elsevier Books, Oxford},\ \bibinfo {year}
  {2002})\BibitemShut{NoStop}%
\bibitem{Privalov93}%
  \BibitemOpen
  \bibfield{author}{%
  \bibinfo {author} {\bibfnamefont{G.}~\bibnamefont{Makhatadze}}\ and\ \bibinfo
  {author} {\bibfnamefont{P.}~\bibnamefont{Privalov}},\ }%
  \bibfield{journal}{%
  \bibinfo {journal} {J.~Mol.\ Biol.}\ }%
  \textbf{\bibinfo {volume} {232}},\ \bibinfo {pages} {639} (\bibinfo {year}
  {1993})\BibitemShut{NoStop}%
\bibitem{1EYDPDB}%
  \BibitemOpen
  \bibfield{author}{%
  \bibinfo {author} {\bibfnamefont{J.}~\bibnamefont{Chen}}, \bibinfo {author}
  {\bibfnamefont{Z.}~\bibnamefont{Lu}}, \bibinfo {author}
  {\bibfnamefont{J.}~\bibnamefont{Sakon}},\ and\ \bibinfo {author}
  {\bibfnamefont{W.}~\bibnamefont{Stites}},\ }%
  \bibfield{journal}{%
  \bibinfo {journal} {J.Mol.Biol.}\ }%
  \textbf{\bibinfo {volume} {303}},\ \bibinfo {pages} {125} (\bibinfo {year}
  {2000})\BibitemShut{NoStop}%
\bibitem{1YMBPDB}%
  \BibitemOpen
  \bibfield{author}{%
  \bibinfo {author} {\bibfnamefont{S.}~\bibnamefont{Evans}}\ and\ \bibinfo
  {author} {\bibfnamefont{G.}~\bibnamefont{Brayer}},\ }%
  \bibfield{journal}{%
  \bibinfo {journal} {J.Mol.Biol.}\ }%
  \textbf{\bibinfo {volume} {213}},\ \bibinfo {pages} {885} (\bibinfo {year}
  {1990})\BibitemShut{NoStop}%
\bibitem{VMD}%
  \BibitemOpen
  \bibfield{author}{%
  \bibinfo {author} {\bibfnamefont{W.}~\bibnamefont{Humphrey}}, \bibinfo
  {author} {\bibfnamefont{A.}~\bibnamefont{Dalke}},\ and\ \bibinfo {author}
  {\bibfnamefont{K.}~\bibnamefont{Schulten}},\ }%
  \bibfield{journal}{%
  \bibinfo {journal} {J. Molec. Graphics}\ }%
  \textbf{\bibinfo {volume} {14}},\ \bibinfo {pages} {33} (\bibinfo {year}
  {1996})\BibitemShut{NoStop}%
\bibitem{SNase79}%
  \BibitemOpen
  \bibfield{author}{%
  \bibinfo {author} {\bibfnamefont{F.~A.}\ \bibnamefont{Cotton}}, \bibinfo
  {author} {\bibfnamefont{J.}~\bibnamefont{Edward E.~Hazen}},\ and\ \bibinfo
  {author} {\bibfnamefont{M.~J.}\ \bibnamefont{Legg}},\ }%
  \bibfield{journal}{%
  \bibinfo {journal} {Proc.\ Natl.\ Acad.\ Sci.\ USA}\ }%
  \textbf{\bibinfo {volume} {76}},\ \bibinfo {pages} {2551} (\bibinfo {year}
  {1979})\BibitemShut{NoStop}%
\bibitem{PDB}%
  \BibitemOpen
  \bibfield{journal}{%
  \bibinfo {journal} {http://www.rcsb.org/}}%
   (\bibinfo {year} {2009})\BibitemShut{NoStop}%
\bibitem{NAMD}%
  \BibitemOpen
  \bibfield{author}{%
  \bibinfo {author} {\bibfnamefont{J.~C.}\ \bibnamefont{Phillips}}, \bibinfo
  {author} {\bibfnamefont{R.}~\bibnamefont{Braun}}, \bibinfo {author}
  {\bibfnamefont{W.}~\bibnamefont{Wang}},\ and\ \bibinfo {author}
  {\bibnamefont{{\it et al}}},\ }%
  \bibfield{journal}{%
  \bibinfo {journal} {J.~Comp.\ Chem.}\ }%
  \textbf{\bibinfo {volume} {26}},\ \bibinfo {pages} {1781} (\bibinfo {year}
  {2005})\BibitemShut{NoStop}%
\bibitem{Russel}%
  \BibitemOpen
  \bibfield{author}{%
  \bibinfo {author} {\bibfnamefont{W.}~\bibnamefont{Russel}}, \bibinfo {author}
  {\bibfnamefont{D.}~\bibnamefont{Saville}},\ and\ \bibinfo {author}
  {\bibfnamefont{W.}~\bibnamefont{Schowalter}},\ }%
  \emph{\bibinfo {title} {Colloidal Dispersions}}\ (\bibinfo {publisher}
  {Cambridge University Press},\ \bibinfo {year} {1989})\BibitemShut{NoStop}%
\bibitem{Mallik02}%
  \BibitemOpen
  \bibfield{author}{%
  \bibinfo {author} {\bibfnamefont{B.}~\bibnamefont{Mallik}}\ and\ \bibinfo
  {author} {\bibfnamefont{T.~L.}\ \bibnamefont{A.~Masunov}},\ }%
  \bibfield{journal}{%
  \bibinfo {journal} {J. Comp. Chem.}\ }%
  \textbf{\bibinfo {volume} {23}},\ \bibinfo {pages} {1090} (\bibinfo {year}
  {2002})\BibitemShut{NoStop}%
\bibitem{Zhou02}%
  \BibitemOpen
  \bibfield{author}{%
  \bibinfo {author} {\bibfnamefont{H.-X.}\ \bibnamefont{Zhou}},\ }%
  \bibfield{journal}{%
  \bibinfo {journal} {BJ}\ }%
  \textbf{\bibinfo {volume} {83}},\ \bibinfo {pages} {2981 – 2986} (\bibinfo
  {year} {2002})\BibitemShut{NoStop}%
\bibitem{Myo04}%
  \BibitemOpen
  \bibfield{author}{%
  \bibinfo {author} {\bibfnamefont{J.~P.}\ \bibnamefont{Collman}}, \bibinfo
  {author} {\bibfnamefont{R.}~\bibnamefont{Boulatov}}, \bibinfo {author}
  {\bibfnamefont{C.~J.}\ \bibnamefont{Sunderland}},\ and\ \bibinfo {author}
  {\bibfnamefont{L.}~\bibnamefont{Fu}},\ }%
  \bibfield{journal}{%
  \bibinfo {journal} {Chem. Rev.}\ }%
  \textbf{\bibinfo {volume} {104}},\ \bibinfo {pages} {561} (\bibinfo {year}
  {2004})\BibitemShut{NoStop}%
\bibitem{Schortle01}%
  \BibitemOpen
  \bibfield{author}{%
  \bibinfo {author} {\bibfnamefont{D.}~\bibnamefont{Shortle}}\ and\ \bibinfo
  {author} {\bibfnamefont{M.~S.}\ \bibnamefont{Ackerman}},\ }%
  \bibfield{journal}{%
  \bibinfo {journal} {Science}\ }%
  \textbf{\bibinfo {volume} {293}},\ \bibinfo {pages} {487 } (\bibinfo {year}
  {2001})\BibitemShut{NoStop}%
\end{thebibliography}

%

\end{document}